**Thermalization of fluorescent protein exciton-polaritons at room temperature**

*Sitakanta Satapathy[1], Bin Liu[3], Prathmesh Deshmukh[1,2], Paul M. Molinaro[1,2], Florian Dirnberger[1], Mandeep Khatoniar[1,2], Ronald L. Koder[1,2] and Vinod M. Menon[1,2]\**

[1] Department of Physics, Center for Discovery and Innovation, The City College of New York, 85 St. Nicholas Terrace, New York, NY 10031, USA.

[2] The PhD Program in Physics, The Graduate Center of the City University of New York, 365 5th Ave, New York, NY, 10016, USA.

[3] Department of Electrical Engineering and Computer Science, University of Michigan, Ann Arbor, Michigan 48109, United States.

\* E-mail: vmenon@ccny.cuny.edu

**Abstract:** Fluorescent proteins (FPs) have recently emerged as a serious contender for realizing ultra-low threshold room temperature exciton-polariton condensation and lasing. Our contribution investigates the thermalization of FP microcavity exciton-polaritons upon optical pumping under ambient conditions. We realize polariton cooling using a new FP molecule, called mScarlet, coupled strongly to the optical modes in a Fabry–Pérot cavity. Interestingly, at the threshold excitation energy (fluence) of ~9 nJ/pulse (15.6 mJ/cm$^2$), we observe an effective temperature, $T_{eff}$ ~350 ± 35 K close to the lattice temperature indicative of strongly thermalized exciton-polaritons at equilibrium. This efficient thermalization results from the interplay of radiative pumping facilitated by the energetics of the lower polariton branch and the cavity $Q$-factor. Direct evidence for dramatic switching from an equilibrium state into a metastable state is observed for the organic cavity polariton device at room temperature via deviation from the Maxwell-Boltzmann statistics at $k_∥ = 0$ above the threshold. Thermalized polariton gases in organic systems at equilibrium hold substantial promise for designing room temperature polaritonic circuits, switches, and lattices for analog simulation.

# 1. Introduction

Organic exciton-polaritons are hybrid quasiparticles, resulting from the strong or ultra-strong coupling between electronic excitations of a molecule and resonant photonic modes in a cavity.[1-8] The huge binding energy and oscillator strength of Frenkel excitons in organic molecules can lead to large vacuum Rabi splitting energies for organic polaritons even at room temperature. Hence, they have been the cornerstone of intense research for realizing emergent photonic and optoelectronic phenomena, such as Bose-Einstein like-condensation of polaritons and polariton lasing,[9-24] nonlinear harmonic generation,[25,26] superfluidity,[27,28] all-optical transistor,[15,29,30] polariton LED,[31,32] polariton-mediated energy transfer,[33-36] and polariton-assisted dynamical modification of molecular excitations.[36-40] Among these the phenomema of polariton condensation, where a transition to a macroscopic coherent state occurs above threshold has been the most widely studied and forms the basis of novel polaritonic device applications. Various organic molecules have been employed to demonstrate condensation, for example, fluorescent proteins, [14,17,22,24], organic crystals, [10,41] thin disordered molecular films, [9,11-14,16,19,20,22,23,25-28] and fluorescent polymers. [15,18,21,30]

A central question in the context of polariton condensates addresses their thermalization. In the case of organic polariton systems, thermalization has been studied mostly for plasmonic lattice coupling.[42-44] However, almost all condensation and lasing experiments in organic molecular systems have been demonstrated using Fabry–Pérot cavities with DBR mirrors.[9-24,29,30] A puzzling aspect in the context of both plasmonic lattice systems and microcavity systems is the ability of the polaritons to thermalize despite their short lifetime (on the order of picoseconds or less).[21] This is due to the efficient polariton relaxation via the release of acoustic phonons.[44] Here, we investigate the thermalization of organic polaritons formed in mScarlet, a new class of red fluorescent protein (RFP). We explore a DBR-Ag $\lambda$-cavity system comprising a red fluorescent protein (RFP), mScarlet, as the active exciton layer to achieve strong coupling. The cavity-$Q$ factor and the radiative pumping of the lower polariton branch (LP1) is shown to play an essential role in achieving room temperature equilibration and effective cooling of the organic polariton gas. Microcavity strong coupling of RFPs intermixed with enhanced green fluorescent proteins (eGFPs) were first reported by Dietrich and co-workers.[45] Prior to this, the same group also proposed the idea of introducing eGFP into simple laminated microcavities to achieve room temperature polariton condensation via strong coupling.[17] In contrast to mCherry, which was previously used to realize lattice polariton condensates,[14,22,24] the protein mScarlet is a highly monomeric, β-barrel structure-based RFP with a record 3.5-times higher oscillator strength than mCherry.[46] The organic protein possess a fluorescence lifetime of ~

3.9 ns with a quantum yield of ~ 70% in solid-state. In addition, it is strongly tolerant under high-power laser irradiation and exceptionally robust against exciton-exciton annihilation and bimolecular quenching at higher excitation densities. All these features are promising for polariton condensation. We compare two-cavity samples with different $Q$-factors (160 and 330) under optical pumping at ambient conditions. Our experiments show a considerable nonlinear increase of the photoluminescence intensity, linewidth collapse, and an increase in the relative blueshift of the high-$Q$ cavity sample. Most importantly, we observe the cooling of exciton-polaritons via room temperature thermalization with the equilibrium switching to a metastable state for the high-$Q$ cavity sample at $k_\parallel = 0$. This is attributed to a combined effect resulting from the cavity $Q$-factor and the energetic position of the lower polariton (LP1) branch.

## 2. Results & Discussion

### 2.1. Cavity characterization

A schematic of our device is shown in Figure 1a. The system comprises of a thin film of mScarlet (active layer) FP embedded in a microcavity comprising a distributed Bragg reflector (DBR) at the bottom and a silver (Ag) mirror at the top. The DBR comprises 10.5 alternating pairs of dielectric ($SiO_2/TiO_2$) layers with a stopband centered at 2.00 eV (620 nm) and a reflectivity of > 99.9 % between 1.76 eV (702 nm) and 2.23 eV (555 nm). For comparison, we fabricated two strongly coupled cavity samples with varying cavity detuning of $E_{ex} - E_{cav} = (22 \pm 2)$ and $(45 \pm 3)$ meV by adjusting the mScarlet film thicknesses (Methods). The thickness of the film is tuned to support two photonic modes within the cavity resulting in a low-$Q$ (LQ) cavity with a $Q$-factor of ~ 160 and a high-$Q$ (HQ) cavity with a $Q$-factor of ~ 330. Figure 1b compares the absorption and photoluminescence (PL) spectra of bare mScarlet film on quartz substrate. The absorption exhibits a maximum at 2.16 eV (574 nm) with a weak shoulder at 2.31 eV (537 nm), whereas the emission peaks at 2.05 eV (604 nm). White light reflectivity maps for both cavity samples are given in Figure S1. Figures 1c and 1d show the angle-resolved photoluminescence (PL) spectra for the strongly coupled LQ and HQ cavity samples, respectively. The positions and the curvatures of the experimentally measured polariton dispersions are reproduced using the transfer matrix method and are fitted using a coupled oscillator model. Subsequently, the cavity parameters are estimated from the fits, which gives

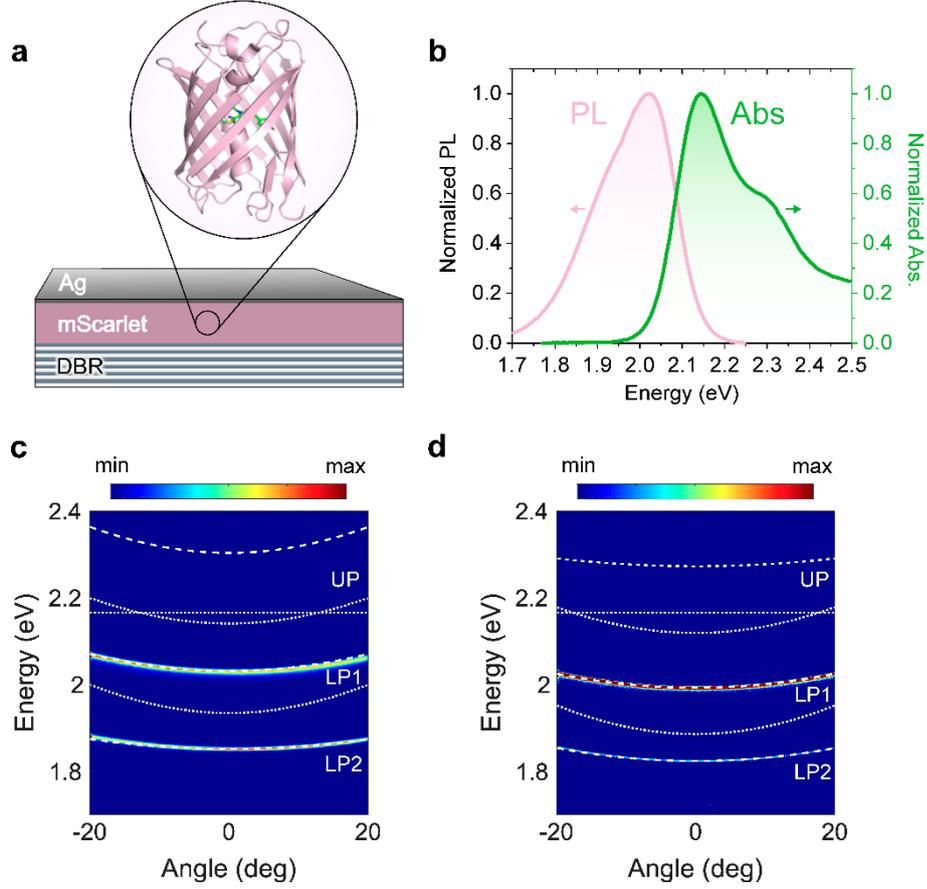

**Figure 1. a)** Schematic of the thin film of mScarlet embedded in a microcavity [Zoom view: schematic image (side view) of the mScarlet molecular structure with the chromophore (green) surrounded by the β-barrel (pink)]. JMol 14.28.3. was used to create the image. **b)** Absorption (green line) and PL emission (pink line) spectra of an mScarlet film on a quartz substrate. Angle-resolved PL maps for the strongly coupled microcavity samples, **c)** LQ and **d)** HQ, showing two lower polariton branches (LP1, LP2) and one upper polariton (UP) branch (white dashed lines). White dotted lines represent the cavity and exciton resonances.

Rabi splitting energies ($\hbar\Omega_R$) of ~258 meV (UP and LP1) and 452 meV (UP and LP2) for the LQ cavity sample. Similarly, Rabi splitting energies ($\hbar\Omega_R$) of ~251 meV (UP and LP1) and 464 meV (UP and LP2) are obtained for the HQ cavity sample, respectively. From the analysis of the Hopfield coefficients (cf. Fig. S2), the LP1 for the LQ cavity sample comprises an exciton fraction of $|X_{k_\parallel=0}|^2 \cong 0.40$ and a photon fraction of $|C_{k_\parallel=0}|^2 \cong 0.60$. On the other hand, the HQ cavity sample comprises an exciton fraction of $|X_{k_\parallel=0}|^2 \cong 0.30$ and a photon fraction of $|C_{k_\parallel=0}|^2 \cong 0.70$. Similarly, the LP2 branches for the LQ and HQ cavity samples are comprised of an exciton fraction of $|X_{k_\parallel=0}|^2 \cong 0.35$ and $|X_{k_\parallel=0}|^2 \cong 0.26$, respectively. On the other hand, the photonic fractions are $|C_{k_\parallel=0}|^2 \cong 0.65$ and $|C_{k_\parallel=0}|^2 \cong 0.74$ for the LP2 branches of the LQ and HQ cavity samples, respectively. For $k_\parallel = 0$ in-plane wave vector, the polariton modes of the LQ cavity are positioned at $E_{UP} = 2.30$ eV, $E_{LP1} = 2.05$ eV, and $E_{LP2} = 1.83$ eV. On the other hand, the polariton modes of the HQ cavity are positioned at at $E_{UP}$ = 2.25 eV, $E_{LP1}$ = 2.00 eV, and $E_{LP2}$ = 1.80 eV.

## 2.2. Excitation power dependence of optical nonlinearities

Next, we study the excitation fluence-dependent PL response of the polariton modes in our strongly coupled cavity samples. In these experiments, the system is excited from the DBR side through off-resonant optical pumping with a photon energy of 2.53 eV (tuned to a reflectivity minimum of the DBR) using excitation pulses of ~80 fs pulse duration and a 12-µm spot diameter with a nearly Gaussian beam profile (Methods).

Figure 2a shows the excitation fluence-dependent PL response of LP1 in the LQ cavity sample integrated around $k = \pm 2°$. The integrated PL intensity shows a sublinear dependence at low excitation fluence followed by an increase in slope for higher excitation fluence. The sublinear power dependence at low excitation fluence is attributed to the sublinear increase in bare exciton PL observed in a mScarlet film on quartz and the comparably low Q-factor of this cavity sample (cf. Figure S4). On the other hand, at higher excitation fluence, partial polariton thermalization gives rise to a change in the slope of the integrated PL intensity (cf. detailed discussion below). While the PL linewidth does not change significantly below an excitation fluence (energy) of 60.7 mJ/cm$^2$ (~35 nJ/pulse), it increases substantially for larger excitation fluence. As shown in Figure 2b, the PL peak shifts towards higher energies below 43.4 mJ/cm$^2$ (~25 nJ/pulse), but decreases with further increase in excitation energy, which is attributed to the onset of photobleaching.

Next, we discuss the excitation fluence-dependent PL response of the HQ cavity sample shown in Figure 2c,d. Interestingly, we observe a superlinear increase of the PL intensity by one order of magnitude around excitation fluences of 15.6 mJ/cm$^2$ (~9 nJ/pulse). This superlinear response accompanied by a characteristic collapse of the linewidth, from 5.9 to 3.4 meV, marks a threshold behavior usually attributed to the stimulated scattering of polaritons into the ground state.[47] The step-like increase in the LP1 energy around excitation fluences of 15.6 mJ/cm$^2$ (9 nJ/pulse) is followed by a change in slope until the onset of photobleaching ~ 60 - 70 mJ/cm$^2$ (35-40 nJ/pulse). The photobleaching is also evident from the linewidth increase (cf. Figure 2c) around the same excitation fluence. We systematically captured this behavior from various areas of the HQ microcavity sample. As discussed in previous reports,[47] the magnitude of blueshift in both the cavity samples is a combined effect resulting from the saturation-induced quenching of the Rabi splitting energy and the cavity mode energy renormalization as a consequence of the change in the effective refractive index ($n_{eff}$) of the cavity from exciting the weakly coupled or uncoupled molecules. On the other hand, a step-like behavior instead of a continuous increase is a characteristic of organic microcavity samples. It is attributed to a competition between the intermolecular energy transfer and stimulated relaxation to the ground state of the polariton.[47,48] The negligible energy shift of the LP1 mode below the threshold is attributed to an intermolecular energy migration process that could rapidly depolarize the fluorescence. [49,50] On the other hand, the change in the slope above the threshold indicates stimulated scattering of the optical dipoles (strongly coupled molecules) aligned with the excitation laser. While the molecular concentration is identical in both cavity samples, the exciton–photon detuning is smaller in the LQ cavity sample. This leads to a higher impact on the cavity mode energy renormalization resulting in comparably greater shift for the LP1 mode of the LQ cavity.[47] The overall change in polariton energy for both cavity samples are relatively small and does not seem to depend on the cavity Q-factor. The angle-resolved PL emission maps in Figure 2e show the polariton distribution in LP1 of the HQ cavity sample for two excitation fluences (above and below the threshold). At higher excitation power above the threshold, we observed a stimulated scattering of polariton density towards lower $k_\parallel$ vectors, close to $k_\parallel = 0$. In contrast, the PL is distributed along the LP1 branch at all powers below the threshold.

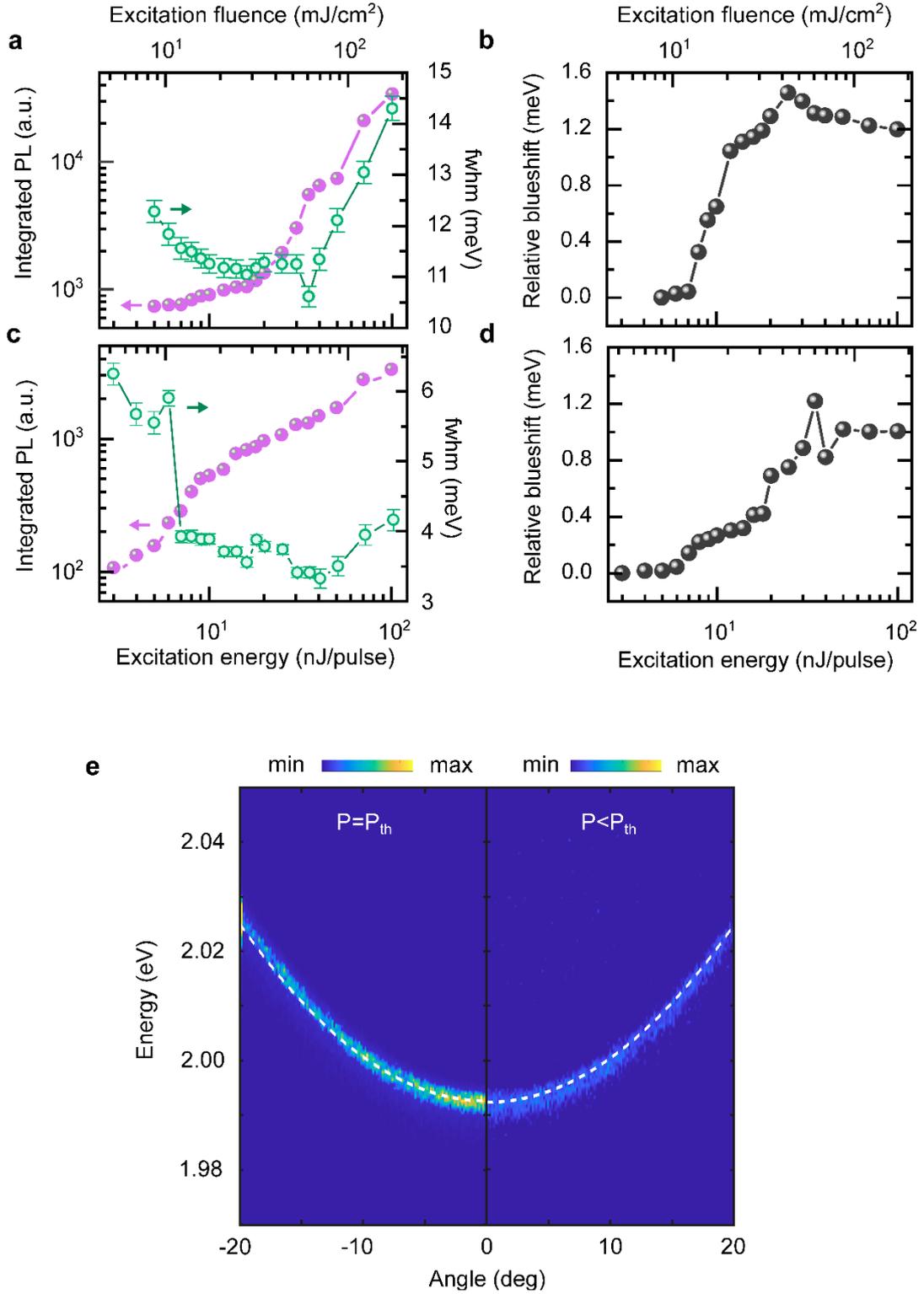

**Figure 2. a)** Integrated PL intensity along with linewidth, and **b)** relative blueshift as a function of different excitation powers for the LQ cavity sample, **c)** Integrated PL intensity along with linewidth, and **d)** relative blueshift as a function of different excitation powers for the HQ cavity sample. The fit results are integrated around $k = \pm 2°$. **e)** Angle-resolved PL maps of the HQ cavity sample below and above the threshold (15.6 mJ/cm$^2$).

Although the angle-resolved emission maps and the power-dependent results indicate the stimulated scattering of polaritons, they are not conclusive evidence for the majority population of polaritons at $k_\parallel = 0$, which is crucial to confirm the condensation of polaritons. Nonetheless, our analysis of the results presented in Figure 2 clearly shows the important role of the cavity *Q*-factor for polariton systems. The signatures of optical nonlinearity and the observed threshold behavior motivate the characterization of polariton thermalization in these microcavity samples.

## 2.3. Thermalization of polaritons

Figure 3 summarizes the results of our thermalization analysis for the LQ and HQ cavity samples. We observed thermalization and cooling with increased pumping for both cavity samples, which is attributed to enhanced polariton scattering at high excitation densities. To extract the effective temperature of the polariton gas, we derive the relative occupation number *N(E)* of the polariton branch by normalizing the PL intensity $I(k_\parallel)$ by the photon fraction at each angle, $|C(k_\parallel)|^2$, giving $N(E) = I(k_\parallel)/|C(k_\parallel)|^2$. The angle-dependence of the energy is given by the corresponding maximum of the PL spectrum. Note that this is a temporally and spatially

integrated measurement of a pulsed driven system which averages over the time evolution during each pulse and the spatial inhomogeneities within the pump spot. Figures 3a and 3b compare the occupation number $N(E)$ of the LP1 branch in both LQ and HQ for different excitation fluences at ambient conditions. Black lines are fits according to a Maxwell-Boltzmann distribution, $N(E) \propto exp(-(E-E_0)/k_B T_{eff})$, from which we extract the effective temperature $T_{eff}$ of polaritons. The result is shown as a function of the excitation fluence in Figures 3c,d for LQ and HQ, respectively. The ground-state energy is determined by $E - E_0 = 0$ meV for $k_\parallel = 0$. For the LQ cavity sample, the polariton occupation of LP1 at low excitation fluence of 8.6 mJ/cm$^2$ (~ 5nJ/pulse) is not thermal, as shown in Figure 3a. On the other hand, we observe a single exponential fit representing the Maxwell-Boltzmann distribution for the excitation fluences ranging from 17.3 mJ/cm$^2$ (~10 nJ/pulse) to 34.7 mJ/cm$^2$ (~20 nJ/pulse). Beyond excitation fluences of 43.4 mJ/cm$^2$ (~25 nJ/pulse), the distribution becomes non-thermal, which is attributed to the onset of photobleaching observed in Figure 2b. For the HQ cavity, below a critical polariton density, the distribution is well fitted by a Maxwell-Boltzmann fit corresponding to a single exponential, i.e., a straight line on a semi-log plot (cf. 12.1 mJ/cm$^2$ and 13.8 mJ/cm$^2$ in Figure 3b). However, at the critical threshold density of 15.6 mJ/cm$^2$ (~9 nJ/pulse), where the ground state of LP1 starts shifting towards higher energy in Figure 2d, we observe an increase of polariton occupation in the ground state at $k_\parallel = 0$. For fluences larger than 15.6 mJ/cm$^2$, the deviation from Maxwell-Boltzmann statistics leads to a significant increase in the polariton distribution at $k_0$ (biexponential behavior). Note that beyond fluences of 60.7 mJ/cm$^2$ (~35 nJ/pulse), the polariton distribution becomes non-thermal with no considerable bi-exponential contribution at $k_0$ (cf. Figure S3). As mentioned above, this is caused by the onset of photobleaching. Unlike in our HQ sample, no increase in $N(E)$ at $k_0$ is noticeable for the LQ sample upon increasing the polariton densities.

Figures 3c and 3d show extracted effective temperatures $T_{eff}$ as a function of excitation fluence for both cavity samples, respectively. The LQ cavity sample shows a decrease from around 540 ± 30 K to the lowest value of 458 ± 22 K. As shown in Figure 4a, although the LP1 branch of LQ cavity is thermalized, the system is characterized by an effective temperature that lies above the lattice temperature of 298 K. As a consequence, polariton emission is observed in a large angle range (inset, Figure 3c). Despite the self-equilibration of polaritons attributed to acoustic phonons,[51] polaritons in the LQ cavity do not reach thermal equilibrium with the lattice and are thus not suited to achieve condensation. Interestingly, for the HQ cavity sample in Figure 3d, the fit values indicate a smooth decrease of $T_{eff}$ from around 515 ± 50 K to 350 ± 35 K around the threshold of 15.6 mJ/cm² (~9 nJ/pulse), which is only slightly above the lattice temperature. Based on the previous reports,[51] it is reasonable to argue that polaritons in HQ cavity sample are thermalized with the lattice and approach thermal equilibrium at this critical threshold. However, upon further increasing the excitation fluence, the HQ cavity sample undergoes a transition to a metastable state where the effective polariton temperature is no longer well defined. The metastable state above the threshold deviates from Maxwell-Boltzmann statistics over a broad range of polariton densities (cf. Figure 3b). This indicates a steady-state quasi-equilibrium condition that predicts a bimodal distribution function for *N(E)* with an increased occupation at $k_\parallel = 0$.[19,20] The calculated lifetime (using exciton and cavity lifetime) of LP1 for the HQ cavity sample is ~ 0.8 ps. On the other hand, vibrational relaxation for such β-barrel fluorescent protein systems occurs within ~ 200 fs,[52], which could further favor the equilibrium dynamics in our cavity polariton device.[44,48] Further evidence for stimulated scattering of LP1 densities towards the ground state ($k_0$, Figure 4b) with increased excitation pump powers can also be observed from the polariton density distribution plot (inset, Fig. 3d) as a function of angle. Note that we did not observe any signatures of optical nonlinearity at the LP2 branch of both the cavity samples despite its relatively high photonic character compared to that at LP1. Also, we didn't observe any evidence of Raman mode assisted decay to the LP modes for any of our cavity samples used in this study (cf. Figure S5, SI).[15,53,54]

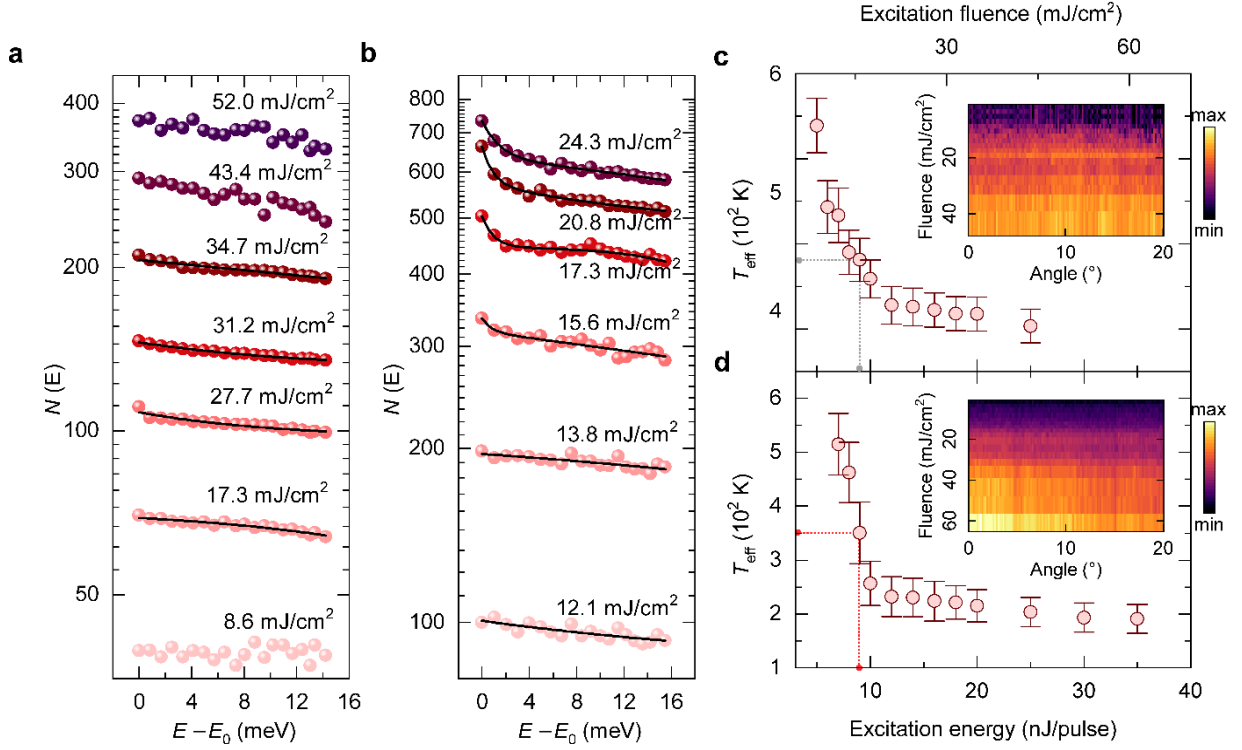

**Figure 3.** Energy distribution of polaritons at $k_\parallel = 0$ for **a)** LQ and **b)** HQ cavity samples for a series of excitation powers below and above threshold (the black lines represent Maxwell-Boltzmann fits). Effective polariton temperature ($T_{eff}$) at $k_0$ as a function of excitation energy/fluence for the **c)** LQ and **d)** HQ cavity samples. The gray dotted line indicates the excitation energy/fluence corresponding to a slope change in $T_{eff}$ for the LQ cavity. The red dotted line indicates the excitation energy/fluence threshold at which the $T_{eff}$ of the HQ cavity system approaches equilibrium. Plots in the inset show maps of the polariton density distribution as a function of the emission angle for different excitation fluences.

From our thermalization analysis, we obtain two important conclusions: First, as shown in Figure 4, the LP1 energy in both the LQ ($E_{LP}$ = 2.05 eV) and HQ ($E_{LP}$ = 2.00 eV) cavity samples is close to the emission maximum of a mScarlet film (2.05 eV). This overlap facilitates the (intracavity) pumping of uncoupled (dark) excitons into the LP1 branch in both cavity samples upon increasing the excitation fluence.[55,56] Second, the larger Q-factor of the HQ sample increases the polariton lifetime and thus results in more strongly pronounced signatures of optical nonlinearities via stronger thermalization and cooling of polaritons within the LP1 branch. On the contrary, the large energy separation between LP2 and the exciton reservoir suppresses efficient scattering and hence no signatures of fluence-dependent optical nonlinearities are observed for LP2.

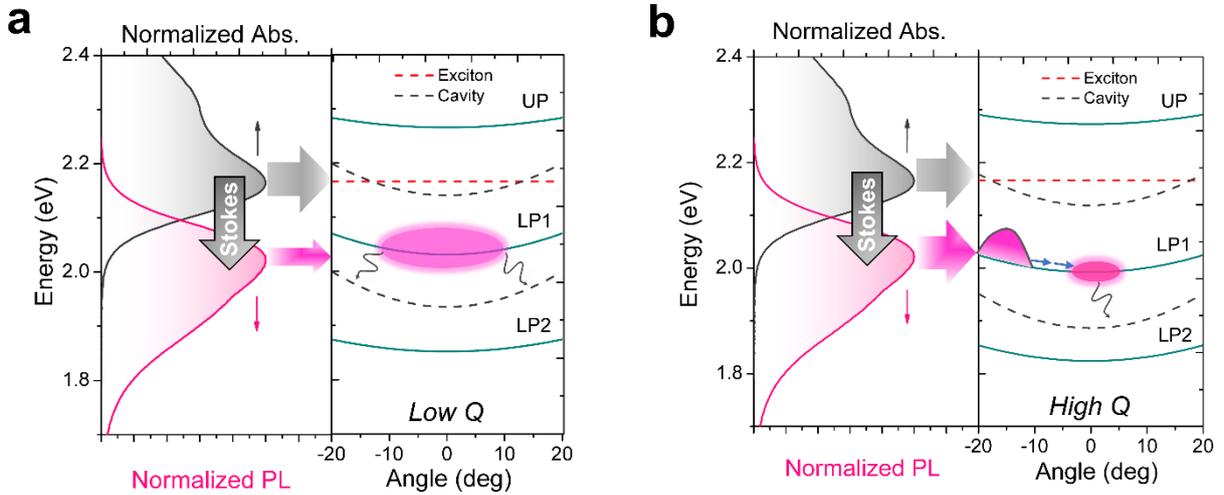

**Figure 4.** Schematic comparing the polariton relaxation and Stokes shift assisted thermalization mechanism across the LP1 branch of the a) LQ and b) HQ cavity samples, respectively.

From the power-dependent experiments and our thermalization analysis, it is clear that both the cavity lifetime and the position of the polariton branch with respect to the emission reservoir play a decisive role in achieving strong thermalization and the cooling of organic exciton-polaritons.[55,56] At larger excitation fluences, the polariton occupation of LP1 in the HQ cavity sample deviates considerably from a Maxwell-Boltzmann distribution, which indicates that the boson statistics of polaritons become more relevant and implies that polariton condensation is approached but not reached (cf. illustration in Figure 4b). The observed increase in the polariton occupation at $k_{\parallel} = 0$ for the high-Q cavity sample is not high enough to induce the thermal phase transition of polaritons into a condensate as observed in the previous reports. [20,22,23]

**3. Discussion and Conclusion**

In summary, we have presented experimental evidence for thermalization and cooling of fluorescent protein (mScarlet) organic exciton-polaritons under ambient conditions using Fabry–Pérot microcavities. A vital result of this study is the observation that the polariton occupation is characterized by a transition from an equilibrium distribution to a metastable state alongside the thermalization and cooling of organic polaritons at room temperature. Except for a higher threshold ~9 nJ pulse$^{-1}$ (15.6 mJ/cm$^2$), the magnitude of change for all the characteristic parameters (integrated emission intensity, linewidth, and blueshift) accounting for optical nonlinearity in our samples are in good agreement with the recent polariton condensation reports obtained from the strongly coupled planar and structured DBR cavities comprising of red and green fluorescent proteins.[14,17,22,24] Our excitation fluence threshold (15.6 mJ/cm$^2$) for stimulated scattering with the use of a DBR-Ag cavity is comparable with some of the reports on BODIPY dye-based polariton condensates in a DBR-DBR cavity.[12,13,57,58] However, the value is higher than some of the other reports on stimulated scattering and polariton lasing of organic systems at room temperature achieved by using the DBR-DBR cavity.[11,48,59] This is attributed to the use of DBR-Ag cavities in our studies, which is intended to simplify the sophisticated fabrication process of a structured top DBR employed previously for such strongly coupled fluorescent protein systems.[60] Although the use of DBR-Ag cavity offers a relatively low $Q$-factor as compared to that obtained from the use of a DBR-DBR or a structured DBR ($Q \sim 10^3$ -$10^4$) cavity for organic molecules,[14,22,24] our simpler cavity devices are attractive considering the well-defined characteristic signatures of optical nonlinearity, stimulated scattering, strongly thermalized polariton relaxation at moderate threshold and operating at room temperature. Our thermalization results hold substantial promise for potential Bose Einstein like-condensation via cooling of polaritons in such fluorescent protein systems using DBR-Ag cavities by tuning the balance between the $Q$-factor and energy position of the polariton mode. Furthermore, patterning of the top silver mirror or the bottom DBR can lead to confined polaritons that will likely have an even lower threshold and help in realizing polariton lattices. We envisage these results to open a new avenue for studying quasi- or non-equilibrium dynamics of organic polaritonic systems and Hamiltonian simulators using room temperature polariton condensates.

## 4. Experimental Section/Methods

*Expression and purification of mScarlet fluorescent protein:* The gene encoding mScarlet with an N-terminal His tag in vector pET24 was the kind gift of Allie Obermeyer of the Department of Chemical Engineering, Columbia University. The vector was transformed into NiCo21 cells and expressed and purified. Briefly, cells were grown in 100mL of LB solution overnight at 37°C, then transferred to 1L of TPP solution and grown at 37° until the solution reached an $OD_{600}$ (Optical density at 600 nm) of roughly 1. Expression of mScarlet was induced by adding 0.5mM IPTG and carried out at 18°C overnight. The cells' harvestation was via centrifugation followed by resuspending in 30 ml 50mM monosodium phosphate, 20mM Imidazole, 300 mM NaCl, pH 8.0 containing 5mg pefabloc and lysed via french press. The lysate was spun down to remove cell debris and added to a column containing HisBind Resin (Novagen), and washed with 100mL of the same buffer. The protein was eluted from the column with buffer 50mM monosodium phosphate, 250mM Imidazole, 300 mM NaCl, pH 8.0, and dialyzed into 20mM sodium bicarbonate buffer at 4°, and lyophilized. Lyophilized mScarlet was resuspended in ultrapure water, and spin concentrated until a concentration of ~250g/L was reached. Protein purity was >95% as determined by SDS-PAGE analysis. Additional water was then added, and the spin concentration was repeated two additional times to remove any residual buffers and salts.

*Fabrication of mScarlet fluorescent protein cavity samples:* The DBR substrates were subjected to O2 plasma treatment for ~ 5 mins, which helped have a better spread for the aqueous mScarlet solution prior to spin coating. The mScarlet (291 mg/mL) solution was then spin-coated at 1500 and 2200 RPMs (acc = 500, t = 120 sec) onto these $O_2$ plasma-treated DBR substrates. The films are subjected to constant pressure (2500 Pascal) treatment in the dark under room temperature for ~48 hrs. This is done to maintain a uniform film thickness throughout a large area range (~2000 microns in XY plane) estimated using the direct method via Profilometer (Bruker Dektak-XT). The thickness values calculated are ~1.6 and ~1.3 μm for spin speeds of 1500 and 2200 rpm, respectively. All these films were then carried out for slow evaporation of Ag (0.2 Å/sec) using the e-beam evaporation technique. We deposited 50 nm of top Ag for LQ cavity, while 100 nm of top Ag was deposited for the HQ cavity. We use two DBR mirrors with center wavelengths at 2.1 eV and 2.0 eV for white light reflectivity and PL measurements from the cavity samples.

*Linearly Optical Spectroscopy*: Spectroscopic ellipsometry (Woollam V-VASE) technique was used to perform the reflection measurements. The measurements are collected at room temperature at an interval of 5°. Solid-state UV-visible measurements on quartz slides are carried out using a Jasco-760 UV-visible spectrophotometer. PL measurements are performed using a homemade setup comprising laser coupled with a Princeton Instruments monochromator with a PIXIS: 256 electron-multiplying charge-coupled device (EMCCD) camera. A 20×, 0.4 numerical aperture (NA) objective was used for all the measurements.

*Pump Power-dependent Photoluminescence Measurements:* The pump beam (490 nm) was generated in a collinear optical parametric amplifier (Light conversion) pumped by the 800 nm output of an amplified Ti: sapphire laser (Coherent Astrella, 1 kHz), and the pump beam was focused onto the sample using the 20x objective (NA=0.45) with a spot size of 12 μm in diameter. Our k-space setup (with the kHz repetition rate laser) designed for the power-dependent PL experiments is aligned and calibrated with this microscope objective (20x, 0.45 NA) to monitor the collapse of PL intensity around the lower k vectors. Photoluminescence emission was collected in reflection configuration using the spectrometer (Princeton Instruments, Acton SpectraPro SP-2500) and charge-coupled device (CCD) camera (Princeton Instruments, PIX 1024B). The residual excitation beam was blocked using a 550 nm long-pass filter. The power dependent PL response from the thin film of mScarlet is recorded at 575 nm excitation pump wavelength using a pulsed laser (Toptica, 78 MHz).

**Supporting Information**

Supporting Information is available from the Wiley Online Library or from the author.

Acknowledgments

The authors acknowledge the use of the nanofabrication and imaging facility at ASRC, CUNY. The authors also acknowledge Prof. Matthew Y. Sfeir (ASRC, CUNY) and the Photonics facility at the ASRC, CUNY for providing access to the laser (Coherent Astrella, 1 kHz) for performing the power dependent measurements on the cavity samples.

**Funding:** This work was supported by National Science Foundation, United States (through NSF-QTAQS OMA-1936351 to V.M.M, S.S, P.D, and M.K), QII (through MCB-2025200 to R.L.K and P.M), and infrastructure support from the National Institutes of Health National Center for Research Resources to the City College of New York (through 5G12MD007603-30 to R.L.K and P.M) and Deutsche Forschungsgemeinschaft (DFG, German Research Foundation) through Projektnummer 451072703 to F.D. B.L acknowledges the financial support from the Department of Energy, United States, under DOE-47851-00-01.**Author Contributions:** V.M.M conceived the idea and designed the experiments with S.S., B.L., and P.D. V.M.M. supervised the project. S.S, B.L., and P.D. contributed equally to this work. S.S. did the thin-film optimization, cavity device fabrication, and optical characterization and analyzed the data. S.S., B.L., P.D., and V.M.M. co-wrote the manuscript. B.L. designed the k-space setup. B.L. and S.S. performed the power-dependent PL measurements. P.D. performed the thermalization studies and calculations and analyzed the thermalization results with S.S. R.L.K. and P.M. performed the synthesis and provided the protein (mScarlet) samples. F.D. and M.K. supported the analysis and editing of the manuscript. All authors discussed the results and commented on the manuscript. All authors have given approval to the final version of the manuscript.

**Conflict of Interest**

The authors declare no conflict of interest.

**References**

[1] R. J. Holmes, S. R. Forrest, Org. Elect. 2007, 8, 77; P. Törmä, W. L. Barnes, *Rep. Prog. Phys.* **2014**, *78*, 013901.

[2] T. W. Ebbesen, *Acc. Chem. Res.* **2016,** *49*, 2403.

[3] A. González-Tudela, P. A. Huidobro, L. Martín-Moreno, C. Tejedor, F. J. García-Vidal, *Phys. Rev. Lett.* **2013,** *110,* 126801.

[4] B. Liu, P. Rai, J. Grezmak, R. J. Twieg, K. D. Singer, *Phys. Rev. B* **2015,** *92,* 155301.

[5] S. Kéna-Cohen, S. A. Maier, D. D. C. Bradley, *Adv. Opt. Mater.* **2013**, *1*, 827.

[6] J. Bellessa, C. Bonnand, J. C. Plenet, J. Mugnier, *Phys. Rev. Lett.* **2004,** *93*, 036404.

[7] D. G. Lidzey, D. D. C. Bradley, M. S. Skolnick, T. Virgili, S. Walker, D. M. Whittaker, *Nature* **1998**, *395*, 53.

[8] F. J. Garcia-Vidal, C. Ciuti, T. W. Ebbesen, *Science* **2021,** *373,* eabd0336.

[9] Y. Qu, S. Hou, S. R. Forrest, *ACS Photonics* **2020**, *7*, 867.

[10] S. Kéna-Cohen, S. R. Forrest, *Nat. Photon.* **2010,** *4,* 371.

[11] T. Ishii, K. Miyata, M. Mamada, F. Bencheikh, F. Mathevet, K. Onda, S. Kéna-Cohen, C. Adachi, *Adv. Opt. Mater.* **2021,** *9*, 2102034.

[12] T. Cookson, K. Georgiou, A. Zasedatelev, R. T. Grant, T. Virgili, M. Cavazzini, F. Galeotti, C. Clark, N. G. Berloff, D. G. Lidzey, P. G. Lagoudakis, *Adv. Opt. Mater.* **2017,** *5,* 1700203.
14

# Supporting Information

**Thermalization of fluorescent protein exciton-polaritons at room temperature**

*Sitakanta Satapathy[1], Bin Liu[3], Prathmesh Deshmukh[1,2], Paul M. Molinaro[1,2], Florian Dirnberger[1], Mandeep Khatoniar[1,2], Ronald L. Koder[1,2] and Vinod M. Menon[1,2]\**

[1] Department of Physics, Center for Discovery and Innovation, The City College of New York, 85 St. Nicholas Terrace, New York, NY 10031, USA.

[2] The PhD Program in Physics, The Graduate Center of the City University of New York, 365 5th Ave, New York, NY, 10016, USA.

[3] Department of Electrical Engineering and Computer Science, University of Michigan, Ann Arbor, Michigan 48109, United States.

\* E-mail: vmenon@ccny.cuny.edu

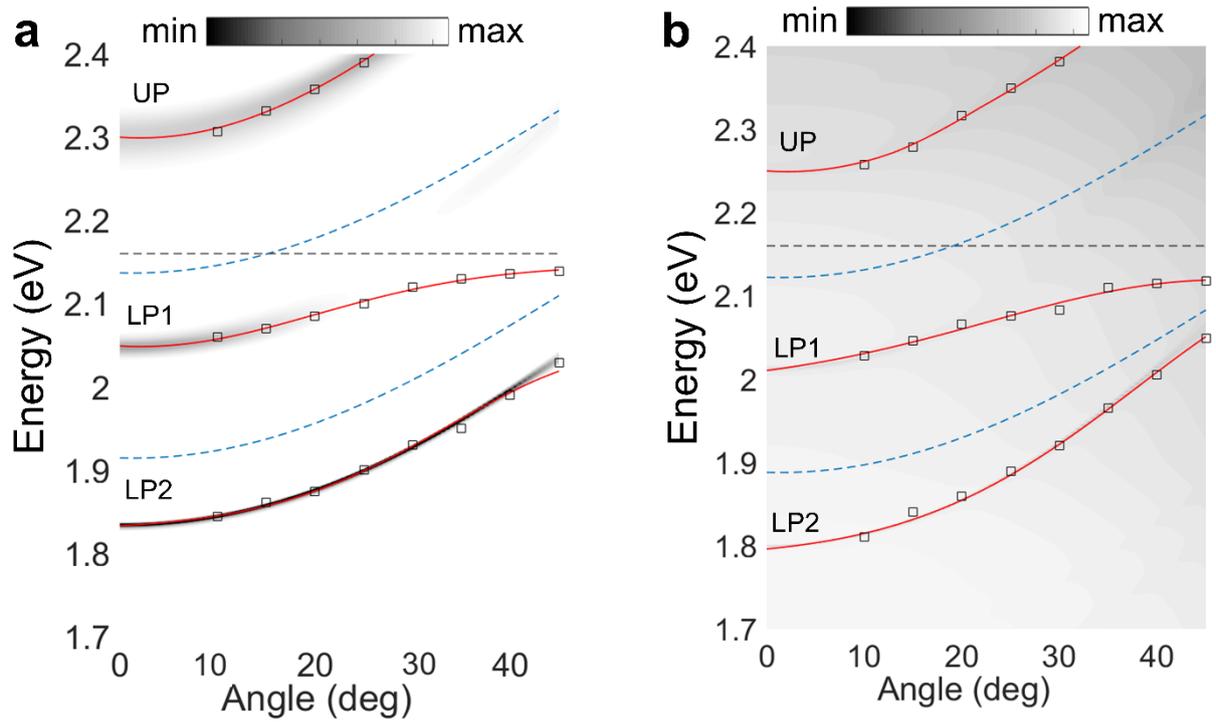

**Figure S1.** Angle-resolved white light reflectivity maps for **a)** LQ and **b)** HQ microcavity samples showing two lower polariton branches (LP1, LP2) and one upper polariton (UP) branch. Black rectangles denote the measured polariton dispersion, on top of the simulation results. The cavity and exciton modes are represented by gray dotted lines and white solid lines, respectively. The white light dispersion from the HQ cavity is weak in contrast due to the high reflectivity of the bottom DBR and top thick Ag. The exciton and cavity modes are represented by gray and blue dashed lines, respectively, while the polariton modes are represented by solid red lines.

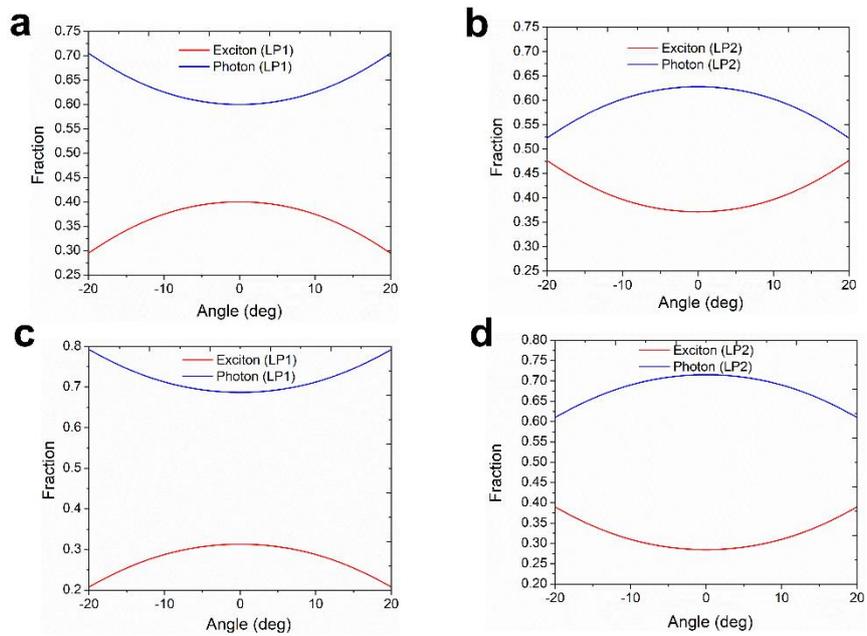

**Figure S2.** Extracted Hopfield coefficients for the lower polariton branches (LP1) and (LP2) for a), b) LQ and c), d) HQ cavity samples.

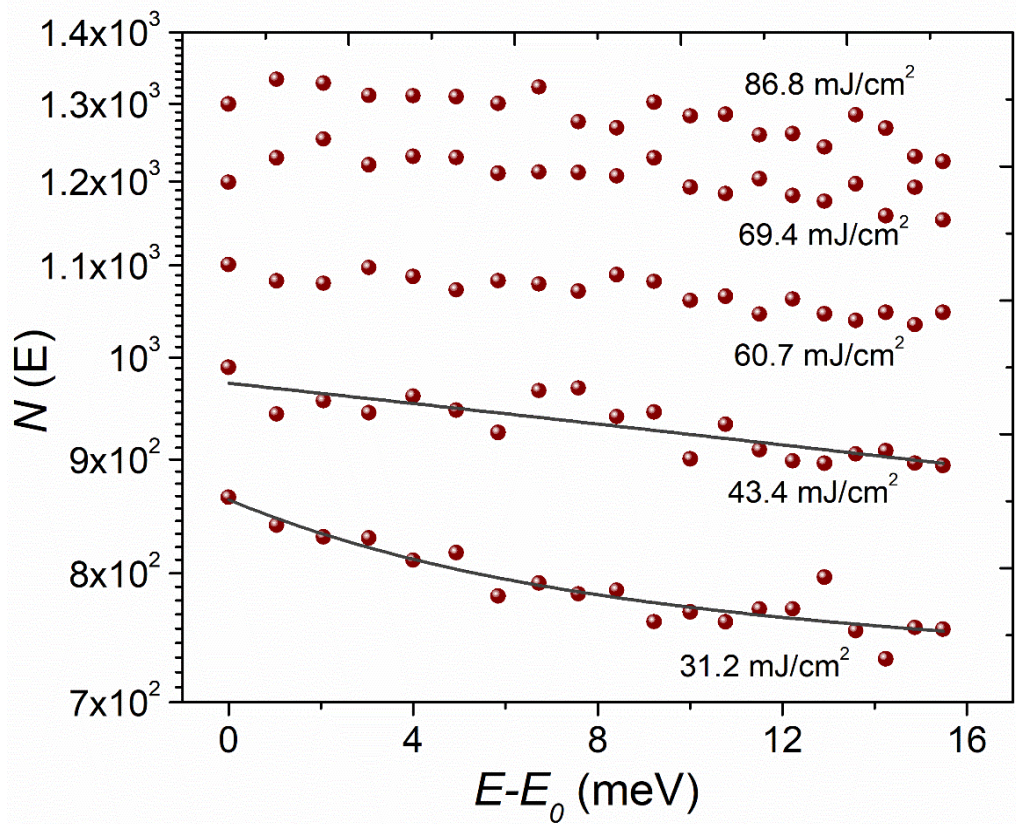

**Figure S3.** Energy distribution of polaritons at $k_{\parallel} = 0$ for the HQ cavity sample for excitation powers high above the threshold (the solid black lines denote the fits).

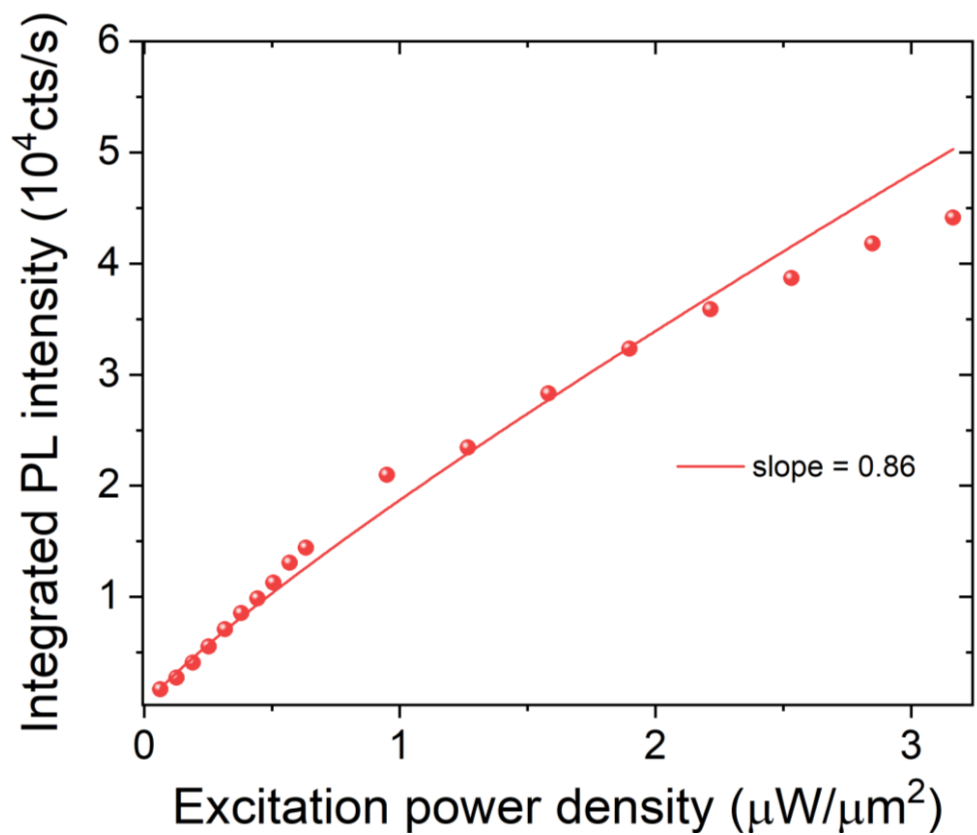

**Figure S4.** Power dependent sublinear photoluminescence response from a thin film of mScarlet upon pulsed laser (575 nm excitation, 78 MHz repetition rate) irradiation.

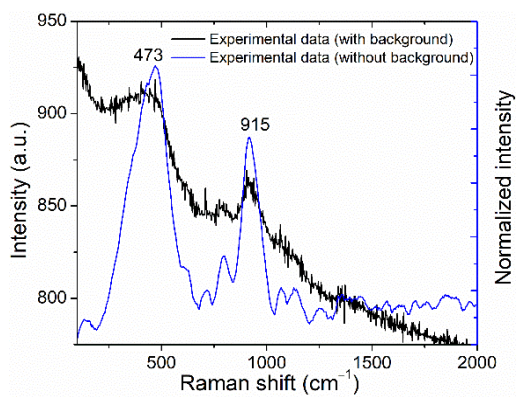

**Figure S5.** Raman spectra for a neat film of mScarlet before (black) and after background PL correction (blue).

Furthermore, stimulated scattering in organic microcavities has also been a consequence of direct vibronic (Raman mode) scattering from the exciton reservoir as a function of the exciton-polariton energy difference. [1,2,3] To verify this hypothesis, we measured the Raman spectra for the neat protein film, as shown in above. From the Raman spectra, we did not observe any significant high energy vibrational modes to be resonant with respect to the exciton-polariton energy difference along with the LP dispersion for any of the cavity samples. Hence, it is difficult to claim the hot exciton scattering mechanism [1] or Raman mode assisted decay for the scattering process. [2,3] Hence, we believe the scattering process in LP1 branch is majorly due to the Stokes-shift assisted radiative pumping mechanism.